\documentclass[twocolumn,showpacs,preprintnumbers,amsmath,amssymb]{revtex4}
\usepackage{graphicx}
\begin{document}
\title{1-D Mott insulator transition of a Bose-Einstein condensate}

\author{R. E. Sapiro, R. Zhang, G. Raithel}

\affiliation{FOCUS Center and Department of Physics, University of
Michigan}

\date{\today}

\begin{abstract}
The superfluid to one-dimensional Mott-insulator transition of a $^{87}$Rb
Bose-Einstein condensate is demonstrated. In the experiment, we apply a
one-dimensional optical lattice, formed by two laser beams with a
wavelength of 852~nm, to a three dimensional BEC in a shallow trap. We use Kapitza-Dirac scattering to determine
the depth of the optical lattice without knowledge of its exact
geometry.  We further study the dynamics
of the transition as well as steady-state phase behavior specific to the
one-dimensional case.
\end{abstract}

\maketitle

Bose-Einstein condensation (BEC) was first demonstrated in
1995~\cite{cornell,ketterle}, creating an explosion of interest in
previously unattainable many-body quantum phenomena.  Of particular
interest is the phase transition from superfluid to Mott
insulator~\cite{greiner}.  In the general, three-dimensional (3-D)
case, this phase transition occurs when a 3-D optical lattice is
applied to a BEC.  As the lattice depth is increased, the BEC
transitions from a superfluid state to a state with a definite
number of atoms in each lattice well. The Mott insulator transition has drawn the
interest of both atomic and condensed matter physicists, due to the
possibilities it creates for simulating ideal, controlled condensed
matter systems. Under the correct circumstances the transition could
be used to create supersolids or other novel phases of
matter~\cite{goral}. Doping the Mott insulator with fermions can be
used to simulate a semiconductor~\cite{ospelkaus}. One can use Feshbach resonances to create
molecules in a Mott insulator with two atoms per
site~\cite{rom}, which could eventually lead to a molecular
BEC~\cite{jaksch,moore}.  The Mott insulator state could also
provide a means to entangle neutral atoms and form a quantum
register for a quantum computer (for a review, see
Ref.~\cite{jakschrev}).  Several laboratories have succeeded in
producing the transition from a BEC to a 3-D Mott
insulator~\cite{greiner,ospelkaus,rom,xu}. In lower dimensions, the
Mott insulator transition has been achieved using
2-D~\cite{phillips,esslinger} and 1-D~\cite{stoferle} Bose gases.
One of the difficulties in attaining the Mott insulator transition
is that the lattice depth necessary to induce the transition
requires high laser power or, more commonly, narrowly focused beams.  Demonstrating the transition from a
BEC to a 1-D Mott insulator (a 3-D BEC in a 1-D lattice formed by two laser beams) is even more difficult because a deeper lattice is needed~\cite{li,oosten}. Furthermore, it can be difficult to determine the exact lattice depth due to uncertainties in the
lattice beam alignment. In this paper, we experimentally
investigate the properties of Mott insulator transition of
a 3-D BEC in a 1-D lattice using a novel, robust method
to calibrate the lattice depth.  In contrast to Ref.~\cite{stoferle}, we apply no transverse lattice potential~\cite{endnote}.  The quantum gas therefore retains superfluidity in the two dimensions transverse to the lattice-beam direction, with observable consequences.

\begin{figure}[t]
\centerline{ \scalebox{.5} {\includegraphics{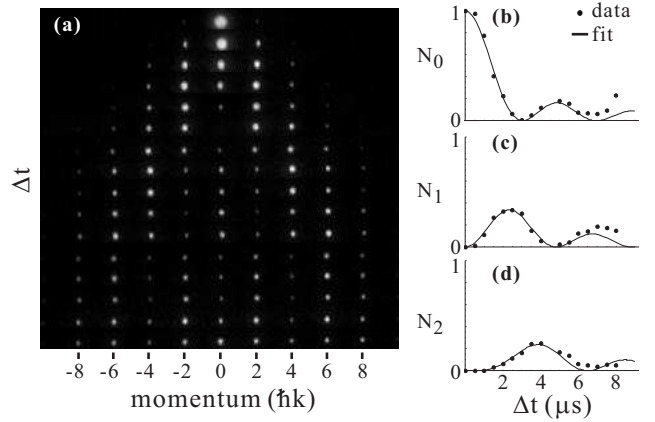}}}
\caption{(a) TOF images of Kapitza-Dirac scattering as a function of
lattice duration $\Delta t$, in steps of $0.5~\mu$s, starting from $0.5~\mu$s.
(b)-(d) Scattering ratios $N_{\rm n}/N_{\rm T}$, as defined in the
text, for the scattering orders $n=0$,1, and 2, respectively, as a
function of lattice duration. The data in (b) are fit with $J_0^2(a
\Delta t)$, with best-fit parameter $a$=0.79. The lines in (b)-(d)
show $J_{\rm n}^2(0.79 \Delta t)$ with respective values of $n$.
\label{kapitzadirac} }
\end{figure}

In the experiment, we start with a $^{87}$Rb
BEC of 5-8$\times 10^4$ atoms in a practically harmonic magnetic trap.  Our BEC apparatus is described
in~\cite{zhang}.  In the data presented, we use magnetic-trap frequencies of 80~Hz and 200~Hz
in the direction of the optical lattice (in
the other two directions, the trap frequencies are 20~Hz and 80~Hz or
40~Hz and 200~Hz, respectively). The optical lattice is formed by a
retro-reflected, far-off-resonance laser beam (wavelength 852~nm, power up to 200~mW after fiber).  The beam is focused into a spot with an intensity full-width half-maximum of 80~$\mu$m.  The depth of the optical lattice is obtained as follows.

When a standing wave of sufficiently short duration is applied to cold atoms such that the atoms are
stationary while the lattice is applied, the system is in the Kapitza-Dirac
scattering regime (analogous to the Raman-Nath regime in optics).  The 1-D optical lattice adds a potential
\begin{equation}
V(x)=-V_0 (1-\cos(2 k_{\rm{L}} x)) \label{potential}
\end{equation}
to the atoms over the time interval $\Delta t$ that the lattice is on. $2V_0$ is the lattice depth, and $k_{\rm{L}}$ is the wavenumber of the lattice beams. Assuming an initial wavefunction $\psi(x,t=0)=1$, the
wavefunction after the lattice pulse is, neglecting a global phase factor,
\begin{align}
\psi(x,t>0)&=\exp\left( i \frac{V_0 \Delta t \cos(2 k_{\rm{L}} x)}{\hbar}\right) \nonumber \\
&=\sum^{\infty}_{n=-\infty} (i)^n J_{\rm{n}}\left(\frac{V_0 \Delta
t}{\hbar}\right)\exp(i 2 n k_{\rm{L}} x)\,\,.  \label{wavefunction}
\end{align}
The expression in the sum shows that the BEC breaks up into momentum components that are integer multiples of $2\hbar k_{\rm{L}}$ with amplitudes given by Bessel functions.
In particular, the order $n=0$ first vanishes at a time $\Delta t_0$ for which
$V_0= 2.4048 \hbar / \Delta t_0$.  The lattice depth $2V_0$ can thus be found by measuring the time $\Delta t$ at which
the $n=0$ order first vanishes. For an atomic polarizability $\alpha$ and a single-beam lattice intensity $I_1$, the lattice depth is also given by $2V_0=\alpha I_1/(2c\epsilon_0)$. Using the above equations, the lattice depth can be experimentally calibrated against arbitrary linear functions of $I_1$, such as the measured beam power. The strength of this method is that no geometrical measurements of the lattice beam size and position are necessary.

To examine the Kapitza-Dirac scattering, the optical lattice is
applied to the BEC for a few microseconds. The lattice and the trap
are then turned off simultaneously. The BEC is allowed to expand
freely for times of flight (TOF) of 16~ms or 12~ms, for the case
of the 80~Hz or 200~Hz traps, respectively. After the expansion, we
take absorption images, shown in Fig.~\ref{kapitzadirac} (a).  Since
the BEC temperature depends on the trap frequency, different trap
frequencies require different TOFs to produce the best image. For
each scattering order $n$, we measure the atom number $N_{\rm n}$
($N_{\rm{T}}$ is the total atom number).  In Fig.~\ref{kapitzadirac}
(b) we plot  $N_0/N_{\rm{T}}$ vs $\Delta t$ and find the lattice
duration $\Delta t_0$ where this ratio first approaches zero; the corresponding lattice depth is $4.8096\hbar/\Delta t_0$.  In this way,
we can determine the proportionality constant relating the
lattice-power reading to the lattice depth. Typically, we find
$\Delta t_0 \sim 3~\mu$s, which is well within the validity range of
the short-pulse approximation underlying the above treatment, as
evidenced by the fits in Fig.~\ref{kapitzadirac} (b) which match the
data well up to 6~$\mu$s. In the measurements presented this paper, this calibration procedure is repeated frequently to account for small changes in the alignment of the lattice beams.  Furthermore, this procedure is used to initially align the lattice with respect to the BEC: better overlap between the lattice and the BEC leads to a smaller $\Delta t_0$.

\begin{figure}[t]
\centerline{ \scalebox{.5} {\includegraphics{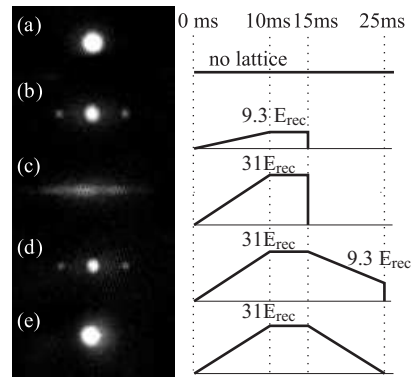}}}
\caption{Left: TOF images. Right: lattice depth as a function of
time for (a) BEC with no lattice, (b) superfluid phase, (c) 1-D Mott
insulator, (d) superfluid phase recovered after Mott insulator, and (e)
BEC with no lattice, recovered after Mott insulator.
\label{becmottbec} }
\end{figure}

With the lattice depth calibrated, we investigate the transition
from a superfluid to a 1-D Mott insulator.  While in Kapitza-Dirac
scattering a sudden, microsecond pulse is applied, the transition of
the BEC from a superfluid to a Mott insulator requires adiabatic
loading of the atoms into the optical lattice, taking of order ten
milliseconds.  We use the amplitude modulation of an AOM to control
the power of our lattice beam.  We ramp up the power of the lattice
beam from zero to its final value over 10~ms, and then hold it there
for 5~ms.  Then, we simultaneously turn off the lattice and the
magnetic trap, and take a TOF measurement.  As can be seen
in Fig.~\ref{becmottbec} (b), for small lattice depths the BEC is
only slightly modulated by the lattice, corresponding to the
appearance of only two weak side peaks, at $\pm 2\hbar k_{\rm L}$.
As the depth of the lattice is increased, the transition to Mott
insulator occurs. The typical signature of the transition is that
the side peaks disappear and the central peak broadens, reflecting
the momentum distribution of the localized wavefunction in a single
lattice well~\cite{greiner}. Here, we find that the system fully
reaches the 1-D Mott insulator state around 30~$E_{\rm{rec}}$.

The Mott insulator transition is a quantum phase transition, and
thus is reversible; to be certain that we have seen the Mott
transition, as opposed to a lattice-induced heating effect, we must
show that we can reverse it.  To demonstrate this, we ramp the
lattice to 31~$E_{\rm{rec}}$ over 10~ms, hold it there for 5~ms, and
then ramp back down over 10~ms.  As can be seen in
Figs.~\ref{becmottbec} (d) and (e), we obtain a modulated superfluid
and BEC when we ramp down to a weak lattice and no lattice,
respectively.  Thus, the effect we see is fully reversible,
providing strong evidence that it is the 1-D Mott insulator
transition.

In the 1-D Mott-insulating state (Fig.~\ref{becmottbec}~(c)), the
quantum gas loses phase coherence in the direction of the optical
lattice while retaining its superfluidity in the other two
directions.  The 1-D Mott insulator can thus be thought of as a
stack of uncorrelated pancake BECs, each containing $\sim$3000 atoms
under the conditions of Fig.~\ref{becmottbec}~(c). As can be seen in
Fig.~\ref{becmottbec} (c), the 1-D Mott insulator expands much
farther in the direction of insulation than in the directions of
superfluidity. This is largely due to the momentum spread of the
pancake BECs in the lattice-beam direction. Examining the TOF image
in Fig.~\ref{becmottbec} (c), we find a velocity spread of $\Delta
p/m_{\rm{Rb}}=8~\rm{mm}/\rm{s}$. Using the Heisenberg uncertainty
relation, $\Delta x \Delta p \geq \hbar/2$, this corresponds to a
localization $\Delta x=46~\rm{nm}$, or 11$\%$ of the lattice period.
Neglecting mean-field effects and using the fact that the lattice
wells are approximately harmonic near their minima, we find an
oscillation frequency of $2 \pi \times 35$~kHz for a lattice with a
depth of 30~$E_{\rm{rec}}$, and velocity and position uncertainties
of 8.9~mm/s and 41~nm, respectively, for the ground state. These
numbers match the values derived from Fig.~\ref{becmottbec} (c)
quite well, showing that the expansion in the lattice-beam direction
is mostly driven by the kinetic energy of the pancake BECs in the
optical-lattice wells.

A more subtle effect is that in the insulating case the expansion
transverse to the lattice-beam direction is considerably slower than
in the lattice-free BEC: about 1.5~mm/s and 2.5~mm/s, respectively.
We attribute the difference to a variation in the manifestation of
the repulsive mean-field potential (estimated to be $\lesssim 1$~kHz
for our BECs in 200~Hz magnetic traps). Without the lattice, the BEC
expansion is driven by a combination of the mean-field pressure and
the kinetic energy of the BEC in the magnetic trap, leading to a
final expansion speed of about 2.5~mm/s in all directions in
Fig.~\ref{becmottbec}~(a). After application of the deep,
Mott-insulating lattice in Fig.~\ref{becmottbec}~(c), the expansion
is mostly driven by the comparatively high kinetic energy of the BEC
pancakes in the optical-lattice wells, leading to a much faster
expansion in the lattice direction. The faster expansion leads to a
reduction of the time over which a substantial mean-field pressure
exists, leading to a reduced final expansion speed transverse to the
lattice, as observed.

\begin{figure}[t]
\centerline{ \scalebox{.4} {\includegraphics{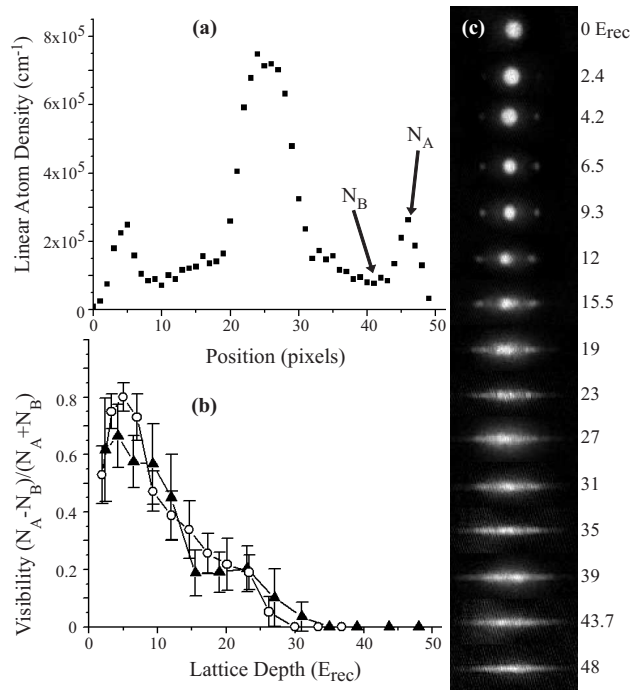}}}
\caption{(a) Linear atom density distribution for a BEC after 12~ms TOF, released from a lattice with a depth of 9.3~$E_{\rm{rec}}$ and magnetic trap with 200~Hz frequency. $N_A$ is the height of
the side peak and $N_B$ is the height of the valley. (b) Visibility as a function of lattice depth from a
80~Hz magnetic trap (circles) and a 200~Hz trap (triangles).  (c) TOF images
as a function of lattice depth for a 200~Hz magnetic trap.
\label{visibility} }
\end{figure}

To quantitatively characterize the 1-D Mott transition, we examine
it as a function of lattice depth.  We use the visibility of the
side peaks, $v$, to map out the transition:
\begin{equation}
v=\frac{N_{\rm A}-N_{\rm B}}{N_{\rm A}+N_{\rm B}} \label{vis}
\end{equation}
where $N_{\rm A}$ is the linear atom density of one side peak, and
$N_{\rm B}$ is the linear atom density at the minimum between the
center peak and the side peak.  The timing of the lattice application is as
described above.  Examining the resulting image, we take the linear
atom density as a function of position in the lattice direction
along the central strip of the image, integrating over three pixels
in height (a pixel in the image corresponds to 6.7~$\mu$m).  For
$N_A$, we choose the local maximum at the side peak, if there is
one, and for $N_B$ we choose the local minimum between the side peak
and the central peak, as shown in the sample data in
Fig.~\ref{visibility} (a).  If there is no local side maximum, we
designate $v=0$.  We calculate the visibility separately for the
side peaks on the left and right, and repeat the calculation for
five separate images at each lattice depth.  We then average the ten
resulting values to get the visibility plotted in
Fig.~\ref{visibility} (b).  As can be seen, the Mott transition
starts around 10~$E_{\rm{rec}}$, where $v$ first reaches a value lower than its initial value. The BEC has fully transitioned
to the Mott insulator state by 30~$E_{\rm{rec}}$, where $v=0$.  Our system contrasts strongly with that of a 1-D Bose gas, where the Mott transition is complete around 10~$E_{\rm{rec}}$~\cite{stoferle}. In 3-D, the Mott transition is complete around
20~$E_{\rm{rec}}$~\cite{greiner}.  This is in general accordance
with the prediction in Ref.~\cite{li} that the 1-D Mott transition
requires a deeper lattice than the 3-D Mott transition. The
specifics do not match, however, in that the lattice depth predicted in Ref.~\cite{li} necessary for the Mott transition is upwards of 50~$E_{\rm{rec}}$
for a system like ours with $\sim$3,000 atoms per lattice well.  As
can be seen in Fig.~\ref{visibility} (b), the Mott transition happens under the same
lattice conditions in the 80~Hz trap and the 200~Hz trap.  This
indicates that at both 80~Hz and 200~Hz the trap has no noticeable
effect compared to the lattice.

Ideally, the visibility should be unity until the Mott transition
starts~\cite{ho}.  In the experiment, however, even a minute thermal
background will lower the visibility, and will disproportionately
affect images with lower $N_A$.  As $N_A,\,N_B \rightarrow 0$, this will prevent the visibility in Eq.~\ref{vis} from reaching unity. For these reasons,
the first few data points in Fig.~\ref{visibility} (b) not only fail
to approach unity, but are even lower than visibilities seen for
lattice depths $\sim5~E_{\rm{rec}}$.  We are, however, confident that
these issues do not affect our determination of where the Mott
insulator transition occurs, because the transition happens at a
lattice depth where both $N_A$ and $N_B$ are large.

\begin{figure}[t]
\centerline{ \scalebox{.35} {\includegraphics{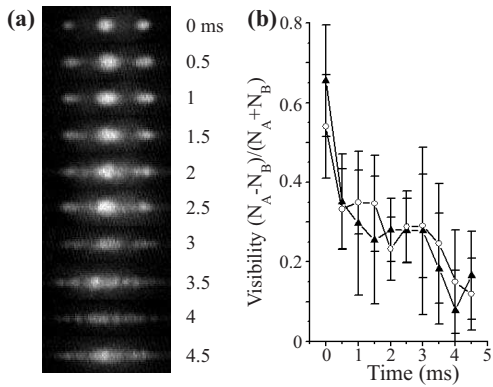}}}
\caption{(a) TOF images as a function of holding time in a lattice of 23~$E_{\rm{rec}}$ depth in a 200~Hz magnetic trap.  (b) Visibility as a function of holding time of the lattice, for an 80~Hz magnetic trap (circles) and a 200~Hz trap (triangles).
\label{dephasing} }
\end{figure}

When the lattice is ramped up, the pancake BECs in the individual lattice wells maintain a global phase until the Mott insulator transition is reached.  Even beyond that point, the pancake BECs in their separate wells require a certain time to entirely dephase with respect to each other; only after the dephasing can one see the typical signatures of the Mott insulator transition (for example, in Fig.~\ref{visibility} (c) for lattices of depths larger than about 20~$E_{\rm{rec}}$).  For deeper lattices, the time needed to dephase is too short to observe.  Within the range of the superfluid to 1-D Mott insulator transition, however, the dephasing time can become sufficiently long to be observable. To measure the dephasing time, we start by ramping the lattice over 10~ms to 23~$E_{\rm{rec}}$, leave the lattice on for a variable hold time, and take TOF images as a function of the hold time, as shown in Fig.~\ref{dephasing} (a).  The visibilities in Fig.~\ref{dephasing} (b) are obtained as described above.  In the case of Fig.~\ref{dephasing}, the dephasing takes of order three milliseconds.

\begin{figure}[t]
\centerline{ \scalebox{.5} {\includegraphics{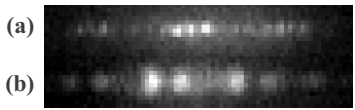}}}
\caption{TOF images of Mott insulators from magnetic traps with frequencies
(a) 80~Hz and (b) 200~Hz.  The speckles vary from shot to shot, and are due to
interference during TOF of pancake BECs from different lattice wells.
\label{interference} }
\end{figure}

In many of the images of the Mott insulator, bright and dark
vertical stripes appear.  These stripes can be seen in nearly all of
the images of the Mott insulator from the 80~Hz trap, and several,
but not all, of the Mott insulator images from the 200~Hz trap. For
example, in Fig.~\ref{visibility} (c), the stripes are visible in the
images at 23 and 35~$E_{\rm{rec}}$. The arrangement of the stripes
appears random, with no repetition of the pattern from image to
image, but the characteristic size of the stripes remains the same
for a given magnetic trap frequency.  We believe that these stripes
represent interference between pancake BECs from different lattice
wells during TOF.  In the 1-D Mott insulator state, each pancake BEC
still has a definite, but random phase.  During TOF, the pancakes
all expand into each other, leading to a characteristic interference
speckle size.  In Fig.~\ref{interference} we show interference
patterns in TOF images for the two different magnetic trap depths.
To find the characteristic speckle size, $\Delta s$, in these
images, we take fast Fourier transforms (FFT) of five images and average
them for each trap depth.  The value of $\Delta s$ is given by the inverse spatial frequency where the FFT signal reaches the noise floor. We find $\Delta s=17~\mu$m for the 80~Hz
trap and 27~$\mu$m for the 200~Hz trap. Using a straightforward
analysis, we estimate that the number of interfering pancakes, $P$,
is related to $\Delta s$ and the TOF, $T$, via $P=2h T/(m\,\Delta
s\,\lambda)$. From this we find $P=20$ and 10 pancakes for the 80~Hz
and 200~Hz traps, respectively.  These numbers agree reasonably well
with our measurements of the size of the BEC in the respective
magnetic traps.

In conclusion, we have demonstrated the 1-D Mott insulator transition by applying a 1-D optical lattice to a BEC.  We used Kapitza-Dirac scattering as an accurate way to calibrate our lattice depth.  We examined time-of-flight images of the BEC as a function of lattice depth, and found that the BEC is fully in the Mott insulator state at 30~$E_{\rm{rec}}$, and requires a certain time to dephase.  Furthermore, we observe random interference patterns in time-of-flight images, providing evidence that the 1-D Mott insulator consists of pancake BECs that still retain superfluid properties, including a definite phase.

We acknowledge the support of AFOSR grant FA9550-07-1-0412 and FOCUS (NSF grant
PHY-0114336), as well as helpful discussions with Professor Luming
Duan.


\begin{thebibliography}{99}

\bibitem{cornell}
M. H. Anderson {\it et al}., 
Science {\bf 269}, 198 (1995).

\bibitem{ketterle}
K.B. Davis {\it et al}., 
Phys. Rev. Lett. {\bf 75}, 3969 (1995).

\bibitem{greiner}
M. Greiner {\it et al}., 
Nature {\bf 415}, 39 (2002).

\bibitem{goral}
K. Goral, L. Santos, and M. Lewenstein, Phys. Rev. Lett. {\bf 88},
170406 (2002).

\bibitem{ospelkaus}
S. Ospelkaus {\it et al}., 
Phys. Rev. Lett. {\bf 96}, 180403 (2006).

\bibitem{rom}
T. Rom {\it et al}., 
Phys. Rev. Lett. {\bf 93}, 073002 (2004).

\bibitem{jaksch}
D. Jaksch {\it et al}., 
Phys. Rev. Lett. {\bf 89}, 040402 (2002).

\bibitem{moore}
M. G. Moore and H. R. Sadeghpour, Phys. Rev. A {\bf 67}, 041603(R)
(2003).

\bibitem{jakschrev}
D. Jaksch, Contemporary Physics {\bf 45} 367-381 (2004)

\bibitem{xu}
K. Xu {\it et al}., 
Phys. Rev. A {\bf 72}, 043604 (2005).

\bibitem{phillips}
I. B. Spielman, W.D. Phillips, and J.V. Porto, Phys. Rev. Lett. {\bf
98}, 080404 (2007).

\bibitem{esslinger}
M. Kohl {\it et al}., 
J. Low Temp. Phys. {\bf 138}, 635 (2005).

\bibitem{stoferle}
T. Stoferle {\it et al}., 
Phys. Rev. Lett. {\bf 92} 130403 (2004).

\bibitem{li}
Jinbin Li {\it et al}., 
arXiv:cond-mat/0311012v2 [cond-mat.soft] 7 Feb 2006.

\bibitem{oosten}
D. van Oosten, P. van der Straten, and H.T.C. Stoof,
Phys. Rev. A {\bf 67}, 033606 (2003).

\bibitem{endnote}
In our work, the parameter $V_{\perp}$ used in Ref.~\cite{stoferle} is $V_{\perp}=0$.

\bibitem{zhang}
R. Zhang, R. E. Sapiro, N. V. Morrow, R. R. Mhaskar, and G. Raithel, submitted.

\bibitem{ho}
Roberto B. Diener {\it et al}., 
Phys. Rev. Lett. {\bf 98} 180404 (2007)

\end{thebibliography}
\end{document}